\documentclass[preprint,amsmath,amssymb]{revtex4}
\usepackage{graphicx}
\usepackage{dcolumn}
\usepackage{bm}
\newif\ifappendixon
\begin{document}


\title{Driving the slow-light soliton by controlling laser field}


\author{A.V. Rybin }
\affiliation{Department of Physics, University of Jyv\"askyl\"a PO
Box 35, FIN-40351 Jyv\"askyl\"a, Finland}
\email{andrei.rybin@phys.jyu.fi}

\author{ I.P. Vadeiko}
\affiliation{School of Physics and Astronomy, University of St
Andrews, North Haugh, St Andrews, KY16 9SS, Scotland}
\email{iv3@st-andrews.ac.uk}

\author{ A. R. Bishop}
\affiliation{Theoretical Division and Center for Nonlinear Studies,
Los Alamos National Laboratory, Los Alamos, New Mexico 87545, USA}
\email{arb@lanl.gov}

\begin{abstract}
In the framework of the nonlinear $\Lambda$-model we investigate
propagation of a slow-light soliton in atomic vapors and
Bose-Einstein condensates. The velocity of the slow-light soliton is
controlled by a time-dependent background field  created by a
controlling laser. For a fairly arbitrary time dependence of the
field we find the dynamics of the slow-light soliton inside the
medium. We provide an analytical description for the nonlinear
dependence of the velocity of the signal on the controlling field.
If the background field is turned off at some moment of time, the
signal stops. We find the location and shape of the spatially
localized memory bit imprinted into the medium. We show that the
process of writing optical information can be described in terms of
scattering data for the underlying scattering problem.
\end{abstract}
\pacs{03.75.Kk, 03.75.Lm, 05.45.-a}
\keywords{lambda model, slow light, soliton}
\maketitle

The nonlinear theory of the Lambda-type model of alkali atoms has
received a new impetus for further development due to significant
progress in the experiments on coherent control of the light-matter
interaction. The experiments on hot and cold atomic vapors
\cite{Hau:1999, Liu:2001, Phillips:2001, Bajcsy:2003, Braje:2003,
Mikhailov:2004, Dutton:2004} demonstrated intriguing possibilities
for realization of nonlinear control over slow-light pulses. From
the theoretical point of view, the most important problem is to
describe the processes of storing and reading of the optical
information. These processes are facilitated by the interaction of
light with the medium and are typically controlled via some
classical external field. In the linear regime the classical field
is assumed to be stronger than the localized pulses of light
carrying the information. However, with modern experimental
developments it is now evident that for an adequate description of
manipulating, storing and reading an optical signal a nonlinear
description becomes necessary. In this paper we develop a general
nonlinear theory of the control over the dynamics of a slow-light
soliton in atomic vapors. In contrast to the linear EIT theory, in
our nonlinear approach the controlling field is allowed to change in
time in a quite arbitrary way and even to vanish.
\begin{figure}
\includegraphics[width=35mm]{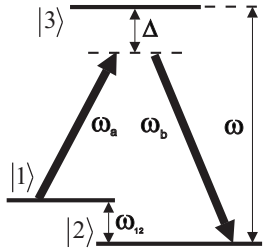}
\caption{\label{fig:spec1} The  $\Lambda$-scheme for working energy
levels of sodium atoms.}
\end{figure}

The working energy levels of alkali atoms are well approximated by
the three-level $\Lambda$-scheme. The structure of levels is given
in Fig.~\ref{fig:spec1}, where $\Delta$ is the detuning of the
carrying frequency from the resonance. The medium is described by
the $3\times3$ density matrix $\rho$ in the interaction picture. In
order to cancel residual Doppler broadening, two optical beams are
chosen to be co-propagating. The electromagnetic fields are
described by the Rabi-frequencies $\Omega_{a,b}$. The field
$\Omega_a$ corresponds to $\sigma^-$ polarization, while the second
$\Omega_b$ field corresponds to $\sigma^+$ polarization. Within the
slowly varying amplitude and phase approximation (SVEPA), dynamics
of the atom-field system is well approximated by the Maxwell-Bloch
equations \cite{gab, Rybin:2004}. Introducing new variables
$\zeta=(x-x_0)/c$, $\tau=t-(x-x_0)/c$ we can rewrite the system of
equations in the following matrix form:
\begin{eqnarray}\label{Maxwell_Bloch}
\partial_\zeta H_I&=&i\frac {\nu_0}4
\left[{D,\rho}\right],\;D=\left(%
\begin{array}{ccc}
  1 & 0 & 0 \\
  0 & 1 & 0 \\
  0 & 0 & -1 \\
\end{array}%
\right),\nonumber\\
\partial_\tau \rho&=&i\left[{\frac\Delta2 D-
    H_I,\rho}\right].
\end{eqnarray}
The parameter $\nu_0$ is the coupling constant. The matrix
$H_I=-\frac12 \left({\Omega_a |3\rangle\langle1|
+\Omega_b|3\rangle\langle2|}\right) +h.c.$ represents the
interaction Hamiltonian.

The system of equations Eqs.(\ref{Maxwell_Bloch}) is exactly
solvable in the framework of the inverse scattering (IS) method
\cite{fad,Hioe:1994, Grobe:1994, Park:1998,gab, Rybin:2004}. This
means that the system of equations Eqs.(\ref{Maxwell_Bloch})
constitutes a compatibility condition for a certain linear system,
namely
\begin{eqnarray}
\label{lin_system}
 &\partial_\tau \Psi= U(\lambda)\,\Psi=\frac i2\lambda D \,\Psi\,  - i \bar H_I
 \,\Psi,\label{lin_system_1}\\
& \partial_\zeta \Psi=
V(\lambda)\,\Psi=\frac{i}{2}\frac{\nu_0\bar\rho}{\lambda-\Delta}
\,\Psi\,.\label{lin_system_2}
\end{eqnarray}
Here, $\lambda\in {\Bbb C}$ is the spectral parameter.

We first described the state of the physical system before the
soliton has entered the medium. In the absence of the soliton the
atoms are assumed to be in the state $|1\rangle$. Notice that this
state is a dark-state for the controlling field $\Omega(\tau)$,
which means that the atoms do not interact with the field
$\Omega(\tau)$ created by the auxiliary laser.  We thus build a
single-soliton solution on the background of the following state of
the overall atom-field system:
\begin{equation}\label{init_fields0}
    \Omega_a=0,\; \Omega_b=\Omega(\tau),\;
    |\psi_{at}\rangle=|1\rangle.
\end{equation}
This configuration corresponds to a typical  experimental setup (see
e.g. \cite{Hau:1999,Liu:2001,Bajcsy:2003}).

The state Eq.(\ref{init_fields0}) satisfies the Maxwell-Bloch
equations Eqs.(\ref{Maxwell_Bloch}). Using the methods of our
previous works \cite{Rybin:2004, ryb7, ryb8}, we construct the
single-soliton solution corresponding to the background field
$\Omega(\tau)$ in the form
\begin{eqnarray}\label{fields_tilde}
 \Omega_a&=&\frac{(\lambda^*-\lambda)w(\tau,\lambda)
} {\sqrt{1+|w(\tau,\lambda)|^2}}\; e^{i \theta_s}\, \mathrm{sech} \phi_s,\\
 \Omega_b&=&\frac{(\lambda-\lambda^*)w(\tau,\lambda) }
{1+|w(\tau,\lambda)|^2}\,e^{ \phi_s}\, \mathrm{sech}
\phi_s-\Omega(\tau),\nonumber
\end{eqnarray}
with the atomic state $ \rho=| \psi_{at}\rangle\langle
 \psi_{at}|$, where
\begin{eqnarray}\label{atoms_tilde}
| \psi_{at}\rangle&=&\frac{\mathrm{Re}\lambda-\Delta-i
\mathrm{Im}\lambda\tanh \phi_s} {|\lambda-\Delta|}  |1\rangle +
\nonumber\\
&&\frac{ \Omega_a}{2|\lambda-\Delta|w(\tau,\lambda)} |2\rangle-
\frac{ \Omega_a}{2|\lambda-\Delta|} |3\rangle.\quad
\end{eqnarray}
Here,
\begin{eqnarray}\label{params_tilde}
 \phi_s&=&
\frac{\nu_0\zeta}{2}\mathrm{Im}\frac1{\lambda-\Delta}+
\mathrm{Re}(z(\tau,\lambda))+\ln\sqrt{\frac{1+|w(\tau,\lambda)|^2}
{1+|w(0,\lambda)|^2}},\nonumber\\
 \theta_s&=&-\frac{\nu_0\zeta}2 \mathrm{Re}
\frac1{\lambda-\Delta}+\mathrm{Im}(z(\tau,\lambda)),\nonumber
\end{eqnarray}
\noindent while the functions $w(\tau,\lambda)$ and
$z(\tau,\lambda)$ are defined below.

In this report we envisage the following dynamics scenario. We
assume that the slow-light soliton   is propagating in nonlinear
superposition with the background field, which is constant at
$\tau\to-\infty$ and vanishes at $\tau\to +\infty$. The speed of the
slow-light soliton is controlled by the intensity of the background
field. Therefore, when the background field decreases, the
slow-light soliton  slows down and stops, eventually disappearing
and leaving behind a standing localized polarization flip, i.e.
optical memory bit. Should the background field increase, the
soliton will emerge again and accelerate accordingly.

To be specific, we define the asymptotic behavior for the field
$\Omega(\tau)$ in the form
\begin{equation}\label{field_asym}
    \Omega(\tau\rightarrow-\infty)=\Omega_0,\;
    \Omega(\tau\rightarrow+\infty)=0.
\end{equation}
The  asymptotic boundary conditions Eq.(\ref{field_asym}) dictate
the following asymptotic behavior for the functions
$w(\tau,\lambda)$ and $z(\tau,\lambda)$:
\begin{eqnarray}\label{w_asym}
w(-\infty,\lambda)&=& w_0 =\frac{\Omega_0}{2 k(\lambda)},\;
w(+\infty,\lambda)=0,\\
\label{z_asym}z(-\infty,\lambda)&=& z_0\tau=i\frac
{|\Omega_0|^2}{4k(\lambda)} \tau,
\end{eqnarray}
where $k(\lambda)=(\lambda+\sqrt{\lambda^2+|\Omega_0|^2})/2$. The
function $z(\tau,\lambda)$ satisfying  the asymptotical conditions
Eq.(\ref{z_asym}) reads
$$z(\tau,\lambda)=z_0\tau+
\int\limits_{-\infty}^\infty\left({\frac i2\Omega^*(\tau')
w(\tau',\lambda)-z_0}\right)\Theta(\tau-\tau')d\tau'.$$

The  function $w(\tau,\lambda)$  is defined by the following
relations

\begin{eqnarray}\label{w_solut}
w(\tau,\lambda)&=&i\int\limits_{-\infty}^\infty e^{-i\,k(\tau-s)}
\Theta(\tau-s) \tilde w(s,\lambda)\,ds,\\
 \label{Riccati_asym}
\tilde w(\tau,\lambda)&=&\frac{\Omega(\tau)}2 +\frac1{k^2}
\left({\frac{|\Omega_0|^2}4 k\,w-\frac{\Omega^*(\tau)}2
(k\,w)^2}\right)\!.\quad\end{eqnarray} Here $\Theta(\tau)$ is the
Heaviside step function. We rewrite the relations
Eqs.(\ref{w_solut}),(\ref{Riccati_asym}) in the form of nonlinear
integral equation, viz.
\begin{eqnarray}\label{Riccati_asym1}
&\tilde w(\tau,\lambda)=\frac{\Omega(\tau)}2
+\int\limits_{-\infty}^\infty
e^{-i\,k(\tau-s)} \Theta(\tau-s) \tilde w(s,\lambda)\,ds\nonumber\\
&\cdot\int\limits_{-\infty}^\infty e^{-i\,k(\tau-s)} \Theta(\tau-s)
\left({\frac{\Omega^*(\tau)}2 \tilde
w(s,\lambda)-\frac{|\Omega_0|^2}4 }\right)\,ds.\quad\end{eqnarray}

\noindent Hence, we can construct a solution  $\tilde
w(\tau,\lambda)$ iterating Eq.(\ref{Riccati_asym1}) and starting
iterations from $\tilde w_0(\tau,\lambda)=\frac12\Omega(\tau)$.
 \noindent

\noindent Notice that the last  term in Eq.(\ref{Riccati_asym})
provides a correction of order $k^{-2}$, because the function
$w(\tau,\lambda)$ asymptotically behaves as $1/k$. It is plain to
see that for the constant controlling field this term vanishes.
Assuming the real constant background field $\Omega_0$, the
imaginary spectral parameter $\lambda_0=-i \varepsilon_0$ in the
lower half-plane, and in the simplifying approximation
$\varepsilon_0\gg\Omega_0$, the solution Eq.(\ref{fields_tilde})
immediately reduces to the conventional form of the slow-light
soliton \cite{Rybin:2004}, viz.
\begin{equation}\label{ss_0}
    \Omega_a=-\Omega_0 e^{i\theta_{s0}}\, \mathrm{sech}(\phi_{s0}),\;
    \Omega_b=\Omega_0\tanh(\phi_{s0}),
\end{equation}
where
\begin{eqnarray}\label{params0_tilde}
 \phi_{s0}&=&
\frac{\nu_0\zeta}{2}\mathrm{Im}\frac1{\lambda-\Delta}+
\tau\;\mathrm{Re}(z_0),\nonumber\\
 \theta_{s0}&=&-\frac{\nu_0\zeta}2 \mathrm{Re}
\frac1{\lambda-\Delta}+\tau\;\mathrm{Im}(z_0),
\end{eqnarray}
and $z_0\approx-|\Omega_0|^2/(4\varepsilon_0)$.

For an arbitrary dependence of the background field on the retarded
time $\tau$, the speed of the slow-light soliton can be represented
in the following
 form:
\begin{equation}\label{speed1}
    \frac{v_g}c=\frac{\partial_\tau\, \phi_s}
    {\partial_\tau\, \phi_s-\partial_\zeta\, \phi_s}.
\end{equation}
It can be readily seen that
\begin{equation}\label{dtau_phi}
\frac{\partial\phi_s}{\partial\tau}=\frac{\mathrm{Im}(\lambda)
|w(\tau,\lambda)|^2}{1+|w^(\tau,\lambda)|^2},\;
\frac{\partial\phi_s}{\partial\zeta}=\frac{\nu_0}{2}
\mathrm{Im}\frac1{\lambda-\Delta}.
\end{equation}
We have thus found a general solution for the velocity $v_g$ of the
slow-light soliton propagating on   an arbitrary time-dependent
background field in terms of the function $\tilde w(\tau,\lambda)$
given by Eq.(\ref{Riccati_asym1}). This result provides a new way to
study dynamics of localized optical signals in the nonlinear EIT
systems. It allows to easily suggest different schemes to slow down,
stop, and reaccelerate slow-light solitonic contribution in the
probing pulse. With such techniques one can introduce a concept of
probing different regions of the media by changing the time that the
soliton dwells around a particular location. This time is important
in the problems when the interaction between light and some
impurities inside the EIT medium is weak and requires slowing the
signal down in the vicinity of these impurities in order to gain
more information about the structure of the medium.

We also introduce a notion of the distance ${\cal L}[\Omega]$ that
the slow-light soliton will propagate until it fully stops. This
quantity is important because it describes the location of an
imprinted memory bit. The brackets $[\cdot]$ indicate a functional
dependence of the distance on the controlling field $\Omega(\tau)$.
To begin with we consider the case when the field is  instantly
switched off at the moment $\tau=0$, i.e.
$\Omega(\tau)=\Omega_0\Theta(-\tau)$. Then we easily find the
solution for $w$ and $z$:
$$w(\tau,\lambda)=w_0\left({\Theta(-\tau)+\Theta(\tau)\;
e^{-i\lambda\tau}}\right),\;z(\tau,\lambda)=z_0\Theta(-\tau)\tau.$$
Hence, we can obtain the distance ${\cal L}_0$ that the soliton will
propagate through from the moment $\tau=0$ until its full stop at
$\tau\rightarrow\infty$:
$$
{\cal L}_0=\frac{c|\Delta-\lambda|^2}{\nu_0|\mathrm{Im}(\lambda)|}
\ln\left(1+|w_0|^2\right).
$$
Here we make use of the assumption that $\mathrm{Im}(\lambda)<0$.

Now, we can give the definition of the distance ${\cal L}[\Omega]$
for a generic field $\Omega(\tau)$ satisfying the conditions
Eq.(\ref{field_asym}). It is convenient to define it as a relative
distance, namely the difference between the absolute coordinate of
the stopped signal at the maximum of the signal and the distance
${\cal L}_0$. The relative distance reads:
\begin{eqnarray}\label{delta_L}
  {\cal L}[\Omega]=\frac{2c|\Delta-\lambda|^2}{\nu_0\mathrm{Im}(\lambda)}
 \!\!\int\limits_{-\infty}^\infty  \mathrm{Re}\left({\frac
i2\Omega^*(\tau)
w(\tau,\lambda)-z_0\Theta(-\tau)}\right)d\tau.\nonumber
\end{eqnarray}
Using the representation Eq.(\ref{Riccati_asym}) we find
\begin{widetext}
\begin{eqnarray}\label{delta_L_asym}
{\cal L}[\Omega]=
\frac{2c|\Delta-\lambda|^2}{\nu_0\mathrm{Im}(\lambda)}
\mathrm{Re}\left({\int\limits_{-\infty}^{+\infty}
\int\limits_{-\infty}^{+\infty} e^{-i\,k(\tau-s)}
\Theta(\tau-s)\left({\frac{|\Omega_0|^2}4\Theta(-\tau)-\frac{\Omega^*(\tau)}2
\tilde w(s,\lambda)}\right)ds\,d\tau}\right)_.
\end{eqnarray}
\end{widetext}
If we assume that $\Omega(\tau)$ is a smooth function and substitute
the solution for $\tilde w(\tau,\lambda)$, we find the result in the
form of a series
$${\cal L}[\Omega]=\frac{2c|\Delta-\lambda|^2}{2\nu_0\mathrm{Im}(\lambda)}
\mathrm{Im}\left({\sum_{n=1}^\infty \frac{I_n}{k^n}}\right),$$ where
$I_n[\Omega]$ are regularized Zakharov-Shabat
functionals~\cite{fad}. The first two functionals read
$I_1[\Omega]=-\int_{-\infty}^{\infty}\left({|\Omega(\tau)|^2-|\Omega_0|^2\Theta(-\tau)}
\right) d\tau$,
$I_2[\Omega]=\frac1{2i}\int_{-\infty}^{\infty}(\Omega^*(s)\partial_s
\Omega(s)-\Omega(s)\partial_s \Omega^*(s)) ds$. The other
functionals can be obtained through the iteration procedure
described above. As it is usual for the boundary conditions of the
finite density type,
  $I_1$ is not a proper functional on
 the complex manifold  of  physical observables, in the sense described in~\cite{fad}. In that sense all other
  functionals in the expansion with respect to $k$ are
 proper. It is a  plausible conjecture that the minimum of the
 functional of length, Eq.(\ref{delta_L_asym}), i.e. $\delta{\cal L}[\Omega]/\delta\Omega=0$
 with
 $\delta^2{\cal L}[\Omega]/\delta\Omega^2>0$,
 is achieved when the controlling field is switched off instantly. Therefore it seems intuitively correct that the
 minimum is
 delivered by the the function
 $\Omega_0\Theta(-\tau)$ discussed above. This conjecture is also
 supported by a  physically relevant case discussed in our
 work \cite{ryb7}. In that reference we solve exactly the case when
 the controlling field vanishes exponentially, i.e.
 $\Omega(\tau)=\Omega_0(\Theta(-\tau)+\Theta(\tau)e^{-\alpha\tau})$ with
 $\Theta(0)=\frac{1}{2}$. In this case the minimum of length is
 delivered by a singular limit $\alpha\to\infty$, i.e. in the
 regime of instant switching off the controlling field.

Another important characteristics of the system is the shape of the
imprinted signal. It is easy to show that the width ${\cal W}_0$ of
the imprinted memory bit is not at all sensitive to the functional
form of $\Omega(\tau)$. This width reads
\begin{equation}\label{width_P}
{\cal W}_0= 4c\ln(2+\sqrt3)
    \frac{|\Delta-\lambda|^2}{\nu_0\left|\mathrm{Im}(\lambda)\right|}.
\end{equation}
In other words, this exact result is valid regardless of how rapidly
we switch the background field off. This means that  specification
of $\Omega(\tau)$ only influences the location of the stored signal
and does not influence its shape. This result is strongly supported
by recent experiments \cite{Dutton:2004}. This reference emphasizes
the phenomenological fact that the quality of the storage is not
sensitive to the regime of  switching off  the control laser. Our
exact result Eq.(\ref{width_P}) provides a rigorous prove for this
experimental observation.

{\em Discussion}. In this report we have  investigated a mechanism
of dynamical control  of the slow-light soliton whose group velocity
explicitly depends on the background field. For quite general
background field, we found the location and shape of the memory bit
written into the medium upon stopping the signal. Remarkably, the
width of this spatially localized standing polarization flip is not
sensitive at all to the functional form of the controlling field and
is defined by the parameters of the soliton only. Our results
provide us with a motivation to formulate a concept of a {\em
solitonic probing tool}. Indeed, the mechanism of control discussed
in this report allows slowing down the soliton at exact locations
inside the medium in a completely controllable fashion. Thus, we can
increase the interaction time in the atom-field system at a
predefined depth of the sample. In principle, this idea should allow
investigation of the structure of the medium. For example, media
doped with passive impurities can be investigated. Finally, it is
worth mentioning that from Eq. (\ref{delta_L_asym}) it is quite
evident that the location of the recorded memory bit is defined,
through the trace formulas, by the scattering data of the spectral
problem Eq. (\ref{lin_system_1}) accompanied by the boundary
conditions Eq.(\ref{field_asym}). A detailed investigation of this
problem will be reported in a forthcoming publication.

IV acknowledges the support of the Engineering and Physical Sciences
Research Council, United Kingdom. Work at Los Alamos National
Laboratory is supported by the USDoE.
\bibliography{paper}
\end{document}